\documentclass[12pt,prd,aps,amssymb,amsmath,tightenlines,showkeys]{revtex4}

\usepackage{graphicx}

\newcommand{\half}{\textstyle{\frac{1}{2}}}
\newcommand{\third}{\textstyle{\frac{1}{3}}}
\newcommand{\quarter}{{\textstyle\frac{1}{4}}}
\newcommand{\threequarter}{{\textstyle\frac{3}{4}}}
\newcommand{\cPT}{\ensuremath{\mathcal{PT}}}

\begin{document}

\title{Nonlinear eigenvalue problems for generalized Painlev\'e equations}
\author{Carl~M.~Bender${}^1$}\email{cmb@wustl.edu}
\author{Javad~Komijani${}^2$}\email{javad.komijani@glasgow.ac.uk}
\author{Qing-hai Wang${}^3$}\email{qhwang@nus.edu.sg}

\affiliation{${}^1$Department of Physics, Washington University, St.~Louis, MO
63130, USA\\
${}^2$School of Physics and Astronomy, University of Glasgow, Glasgow G12 8QQ, United Kingdom\\
${}^3$Department of Physics, National University of Singapore, 117542,
Singapore}

\date{\today}

\begin{abstract}
Eigenvalue problems for linear differential equations, such as time-independent
Schr\"odinger equations, can be generalized to eigenvalue problems for nonlinear
differential equations. In the nonlinear context a separatrix plays the role of
an eigenfunction and the initial conditions that give rise to the separatrix
play the role of eigenvalues. Previously studied examples of nonlinear
differential equations that possess discrete eigenvalue spectra are the
first-order equation $y'(x)=\cos[\pi xy(x)]$ and the first, second, and fourth
Painlev\'e transcendents. It is shown here that the differential equations for
the first and second Painlev\'e transcendents can be generalized to large
classes of nonlinear differential equations, all of which have discrete
eigenvalue spectra. The large-eigenvalue behavior is studied in detail, both
analytically and numerically, and remarkable new features, such as hyperfine
splitting of eigenvalues, are described quantitatively.
\end{abstract}

\keywords{Painlev\'e equation, $\cPT$ symmetry, semiclassical analysis, WKB
theory, asymptotic approximation, eigenvalue, separatrix}

\maketitle

\section{Introduction}\label{s1}
This paper applies the concepts of stability and instability to extend the
notion of an eigenvalue problem for a linear differential equation to an
eigenvalue problem for a nonlinear differential equation. The basic idea was
proposed first in Ref.~\cite{r1} and then developed further in
Refs.~\cite{r2,r3}. In this paper we extend these earlier studies to huge
classes of nonlinear differential equations, all of which have infinite discrete
spectra of eigenvalues and we explore the large-eigenvalue (semiclassical)
behavior of these spectra.

The nonlinear differential equations considered in this paper have a common
structure. On the left side of the equation is a first or second derivative of
the dependent variable $y(x)$ and on the right side is a function $F(x,y)$ of
the independent and dependent variables:
\begin{equation}
y'(x)=F[x,y(x)]\quad{\rm or}\quad y''(x)=F[x,y(x)].
\label{E1}
\end{equation}
We begin by finding explicit solutions $y=f(x)$ to the implicit algebraic
equation $F(x,y)=0$. Because $F(x,y)$ is nonlinear in $y$, we may find multiple
solutions to $F(x,y)=0$ and we think of each such solution $f(x)$ as a fixed
point in function space. We then pose a deceptively simple question: If $f'(x)$
vanishes for large $|x|$, do the solutions to the nonlinear differential
equation approach $f(x)$ for large $x$?

The answer to this question is complicated. For some initial conditions the
solution $y(x)$ of the nonlinear equation may well approach a solution $f(x)$ of
the algebraic equation $F(x,y)=0$. Furthermore, if all functions that are near
$f(x)$ eventually approach $f(x)$ asymptotically, then $f(x)$ is a {\it stable}
fixed point. However, while $y(x)$ may get close to $f(x)$, it may then veer
away from $f(x)$ as $|x|$ increases; in this case $f(x)$ is an {\it unstable}
fixed point. Indeed, $y(x)$ may exhibit interesting behavior where it repeatedly
gets close to a fixed-point function $f(x)$ but is unable to approach $f(x)$
asymptotically because $y(x)$ develops movable singularities. (Such
singularities can occur because the differential equation is nonlinear.) In this
case $y(x)$ may eventually approach another function $f(x)$ or else $y(x)$ may
not even approach any function $f(x)$ at all.

In this paper our interest is focused on the isolated and rare initial
conditions for which the solution $y(x)$ is asymptotic to an {\it unstable}
fixed point $f(x)$. In such a case, if the initial condition is slightly
altered, $y(x)$ will no longer approach the function $f(x)$. Thus, nearby
solutions to the nonlinear differential equation diverge away from $y(x)$. A
function $y(x)$ with this property is called a {\it separatrix} solution to the
differential equation, and because this function is unstable relative to small
changes in the initial conditions we think of it as an eigenfunction. Also, we
regard the initial conditions that generate the separatrix as eigenvalues.

Our use of the terminology ``eigenfunctions'' and ``eigenvalues'' requires some
explanation. Let us recall the features of linear-differential-equation
eigenvalue problems. For example, consider the case of a linear time-independent
Schr\"odinger eigenvalue problem on the infinite domain $-\infty<x<\infty$:
\begin{equation}
-y''(x)+V(x)y(x)=Ey(x),\qquad y(\pm\infty)=0.
\label{E2}
\end{equation}
Here, $E$ is the eigenvalue and $y(x)$ is the eigenfunction. We assume that the
potential $V(x)$ has the property that $V(x)\to+\infty$ as $x\to\pm\infty$ so
that the potential confines an infinite number of discrete-energy bound states.
The bound-state eigenfunctions exhibit several characteristic behaviors:
\begin{enumerate}
\item In the classically forbidden regions the
eigenfunctions vanish exponentially as $|x|\to\infty$ and WKB theory gives the
precise asymptotic behavior of $y(x)$ for large $|x|$. For example, for positive
$x$
\begin{equation}
y(x)\sim C[V(x)-E]^{-1/4}\exp\left[-\int^x dt\sqrt{V(t)-E}\right]\quad(x\to+
\infty),
\label{E3}
\end{equation}
where $C$ is a constant.
\item Assuming that $V(x)$ is real so that the Hamiltonian is Hermitian, in the
classically allowed region [where $E>V(x)$] the eigenfunctions $y(x)$ are
oscillatory. The eigenfunction $y_n(x)$ associated with the $n$th eigenvalue
$E_n$ has $n$ nodes between the turning points and the eigenfunctions exhibit
the phenomenon of interlacing.
\item There is an abrupt transition between oscillatory and exponentially
decreasing behavior, which occurs at the turning points. In the semiclassical
(large-eigenvalue) regime this transition is universally governed by an Airy
function.
\item The growth of eigenvalues for large n is algebraic; typically,
$$E_n\sim \alpha n^\beta\quad(n\gg1),$$
where the constants $\alpha$ and $\beta$ are determined by the potential $V(x)$.
These constants may be calculated directly from the leading-order WKB
quantization condition
$$\int_{x_1}^{x_2}dx\sqrt{E-V(x)}\sim\big(n+\half\big)\pi\quad(n\to\infty),$$
where $x_1$ and $x_2$ are the classical turning points. For instance, the
semiclassical approximation to the eigenvalues of the harmonic oscillator $V(x)=
x^2$ is
$$E_n\sim 2n\quad(n\to\infty),$$
and the semiclassical approximation to the eigenvalues of the anharmonic
oscillator $V(x)=x^4$ is
$$E_n\sim\left[3\sqrt{\pi}\Gamma(\threequarter)/\Gamma(\quarter)\right]^{4/3}
n^{4/3}\quad(n\to\infty).$$
\item The eigensolutions are unstable in the sense that if the parameter $E$ in
the differential equation (\ref{E2}) is slightly different from an exact
eigenvalue $E_n$, then it is not possible to satisfy both homogeneous boundary
conditions in (\ref{E2}) [except for the trivial solution $y(x)\equiv0$]. To
be precise, when $E=E_n$, there exists an eigenfunction solution $y_n(x)$ that
vanishes at $\pm\infty$, but if $E=E_n+\epsilon$, where $\epsilon\neq0$ is
arbitrarily small, then $y(-\infty)$ and/or $y(\infty)$ are infinite. 
\end{enumerate}

For the eigenvalue problem (\ref{E2}), we identify $F[x,y(x)]=[V(x)-E]y(x)$.
Thus, the solution to $F(x,y)=0$ is $f(x)=0$. However, when $E$ is not an
eigenvalue, the solution $y(x)$ to the Schr\"odinger equation does not approach
0 as $x\to\infty$; rather it becomes infinite as $x\to\infty$ and/or $x\to-
\infty$. From WKB theory we know that for large positive $x$ the solution $y(x)$
typically increases exponentially,
$$y(x)\sim D[V(x)-E]^{-1/4}\exp\left[+\int^x dt\sqrt{V(t)-E}\right]\quad(x\to+
\infty),$$
where $D$ is a constant. The function $y(x)$ approaches $0$ as $|x|\to\infty$
only for a special discrete set of values of $E$. Since these values are called
eigenvalues and the associated functions $y(x)$ are called eigenfunctions, we
apply these terms to the nonlinear differential equations that are studied in
this paper. The separatrix solutions that we have found are isolated and the
spectrum of eigenvalues (the initial conditions) is discrete. Moreover, as
physicists, we think of eigenvalues as energies and we are interested in the
high-energy (semiclassical) behavior of the eigenspectrum. We will see that like
the linear case the $n$th eigenvalue in the spectrum of a nonlinear differential
equation often grows algebraically with $n$ like $\alpha n^\beta$ as $n\to
\infty$.

\subsection{Examples of first-order nonlinear eigenvalue problems}\label{ss1a}
\begin{figure}[h!]
\begin{center}
\includegraphics[width=0.8\columnwidth]{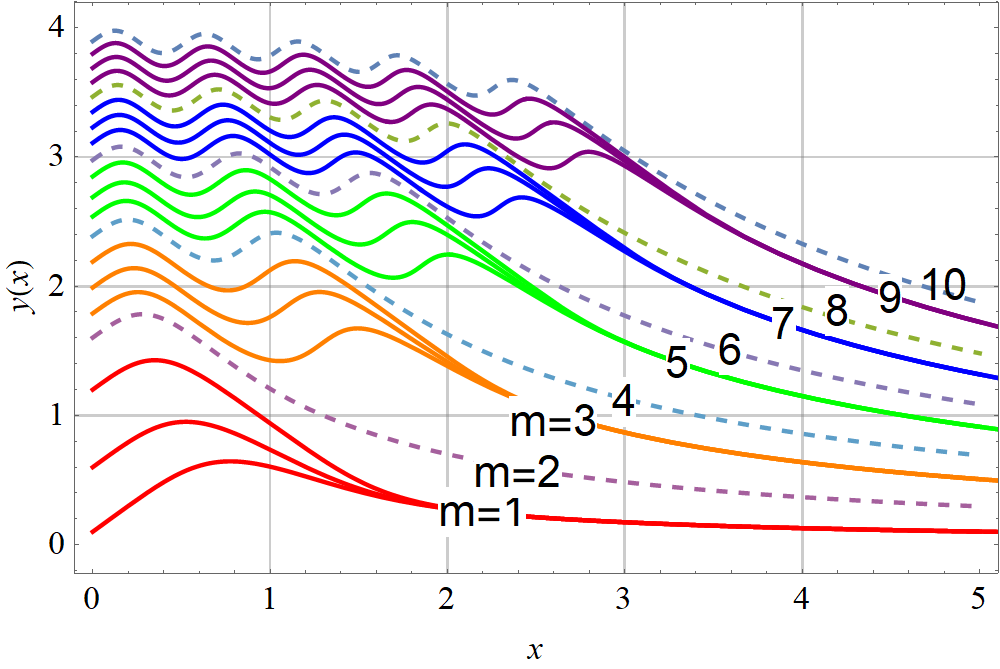}
\caption{\label{F1} Solutions to the differential equation (\ref{E5}) for 20
initial conditions $y(0)$. Observe that the solutions are attracted to the
stable asymptotic behaviors $(m-1/2)/x$ for odd $m$ (solid lines) but veer away
from the unstable asymptotic behaviors $(m-1/2)/x$ for even $m$ (dashed lines).
The dashed lines are separatrix solutions and we think of these separatrices as
eigenfunctions. The $n$th eigenvalue is the value of the $m=2n$th solution
at $x=0$.}
\end{center}
\end{figure}

A toy problem that illustrates the similarities between the properties of
linear eigenvalue problems and nonlinear equations is 
\begin{equation}
y'(x)=\cos[\pi xy(x)],\quad y(0)=E.
\label{E5}
\end{equation}
For this equation $F[x,y(x)]=\cos[\pi xy(x)]$, so $f(x)=(m-\half)/x$ for $m=1,\,
2,\,3,\,\cdots$. Like the solutions in (\ref{E3}), the solutions to this
initial-value problem decay to $0$ monotonically for large $x$. All solutions to
(\ref{E5})
vanish like $\big(m-\half\big)/x$ for large $x$. Twenty such solutions are shown
in Fig.~\ref{F1}; note that the solutions for odd $m$ (solid lines) are stable
[that is, solutions near $f(x)=(m-1/2)/x$ converge to $f(x)$ as $x$ increases].
However, the solutions for even $m$ (dashed lines) are unstable because nearby
solutions diverge away from it. The dashed-line
solutions are separatrices. These solutions have the same features as those
enumerated above for solutions to linear eigenvalue problems, namely,
instability and oscillatory behavior transitioning into monotone decay. Thus, we
call these solutions eigenfunctions and we say that the $n$th eigenvalue $E_n$
is the initial condition that gives rise to the $n$th separatrix: $E_n=y_{2n}(
0)$. Like the eigenvalues associated with linear eigenvalue problems, the
eigenvalues $E_n$ of the nonlinear equation (\ref{E5}) grow algebraically with
large even $m=2n$ \cite{r1,r1a}:
\begin{equation}
E_n\sim2^{5/6}n^{1/2}\quad(n\to\infty).
\label{E6}
\end{equation}

\begin{figure}[h!]
\begin{center}
\includegraphics[width=0.8\columnwidth]{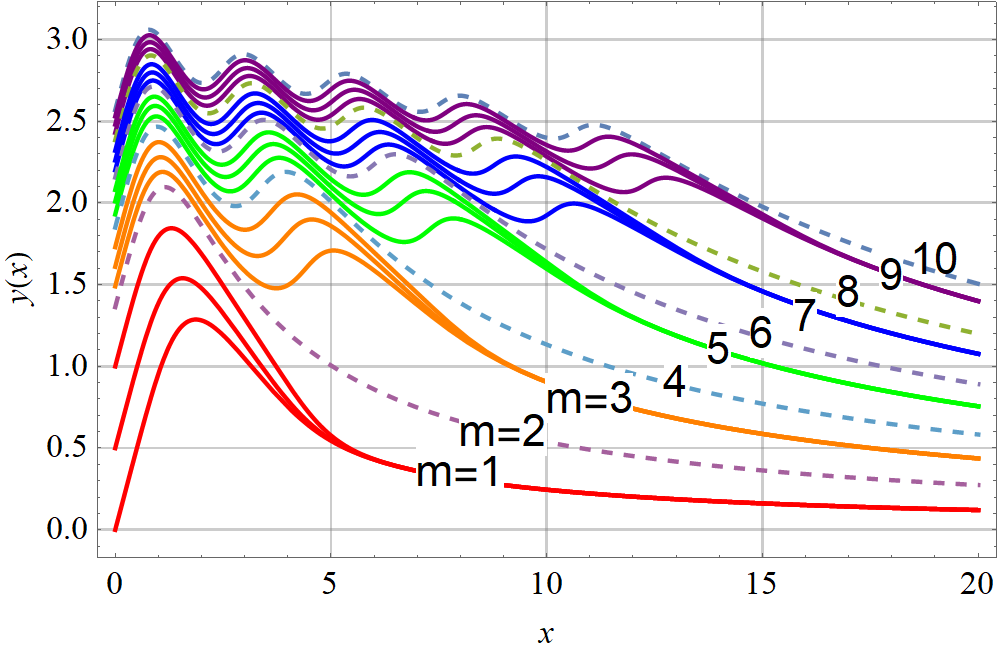}
\caption{\label{F2} Twenty solutions to the nonlinear differential equation
(\ref{E7}). The stable and unstable behavior is qualitatively similar to that
shown in Fig.~\ref{F1}.}
\end{center}
\end{figure}

A slightly fancier first-order nonlinear differential equation is
\begin{equation}
y'(x)={\rm J}_0[xy(x)],\quad y(0)=E,
\label{E7}
\end{equation}
where ${\rm J}_0(x)$ is the Bessel function of order 0. The solutions to this
equation (see Fig.~\ref{F2}) are similar to those of the cosine model in 
Fig.~\ref{F1}. That is, the asymptotic behavior
$$y(x)\sim\omega_m/x,\quad(x\to\infty),$$
where $\omega_n$ ($m=1,2,3,\cdots$) is the $m$th zero of the Bessel function
[${\rm J}_0(\omega_m)=0$], is stable if $m$ is odd (solid lines) and is an
unstable separatrix if $m$ is even (dashed line). We regard the initial value
$y_{2n}(0)$ that leads to the $n$th separatrix (eigensolution) as the $n$th
eigenvalue $E_n$ of this nonlinear differential equation. For large even $m=2n$
we find that
\begin{equation}
E_n\sim\alpha n^{1/4}\quad(n\to\infty).
\end{equation}
Numerical studies suggest that $\alpha$ is close to $\frac{35}{18}$,
which, like the value of $\alpha$ in (\ref{E6}) for the cosine example above,
is slightly less than 2.

\subsection{Eigenvalue problems for Painlev\'e I}\label{ss1b}
The initial-value problem for the first Painlev\'e equation (Painlev\'e I) has
the form
\begin{equation}
y''(x)=6y^2(x)+x,\quad y(0)~{\rm and}~y'(0)~{\rm specified}.
\label{E9}
\end{equation}
(For background information and asymptotic studies of the Painlev\'e equations
see Refs.~\cite{R1,R2,R3,R4,R5,R6,R7,R8,R9,R10,R11}.) For this equation $F(x,y)=
6y^2+x$, and for negative $x$ there are two solutions to $F(x,y)=0$:
$$f_\pm(x)=\pm\sqrt{-x/6}.$$
The solutions to (\ref{E9}) with specified initial conditions at $x=0$ exhibit
three possible behaviors for negative $x$:
\begin{enumerate}
\item After passing through a finite number of double poles on the negative-$x$
axis, the solution $y(x)$ may oscillate about the lower half of the parabola
$f_-(x)$ with decreasing amplitude [see Figs.~1 and 2 (right
panels) in Ref.~\cite{r2}]:
$$y(x)\sim-\sqrt{-x/6}\quad(x\to-\infty).$$
Thus, $f_-(x)$ is a stable fixed point in function space.
\item The solution $y(x)$ may pass through an infinite sequence of double poles
on the negative-real-$x$ axis. Of course, such solutions do not approach a
solution to $F(x,y)=0$ [see Figs.~1 and 2 (left panels) in Ref.~\cite{r2}].
\item After passing through a finite number of poles on the negative-$x$ axis,
the solution $y(x)$ may approach the upper half of the parabola $f_+(x)$
asymptotically [see Figs.~3-7 in Ref.~\cite{r1}]:
$$y(x)\sim\sqrt{-x/6}\quad(x\to-\infty).$$
These solutions are unstable separatrices because nearby solutions veer away
from them and behave like one of the two types of solutions described in 1 and 2
above. We regard these types of solutions as eigenfunctions.
\end{enumerate}

Two types of eigenvalue problems corresponding to two different types of
initial conditions for Painlev\'e I have been studied:
\begin{enumerate}
\item $y(0)=0$: In this case the values of $y'(0)$ that give rise to
separatrices are the eigenvalues. In Ref.~\cite{r2} it was shown that the
large-$n$ behavior of the $n$th eigenvalue is
\begin{equation}
y_n'(0)\sim\pm2\big[\sqrt{3\pi}\,\Gamma\left(\textstyle{\frac{11}{6}}\right)/
\Gamma\left(\third\right)\big]^{3/5}n^{3/5}\quad(n\to\infty).
\label{E10}
\end{equation}
\item $y'(0)=0$: In this case the values of $y(0)$ that give rise to
separatrices are the eigenvalues. Now, the large-eigenvalue behavior is
\begin{equation}
y_n(0)\sim -\big[\sqrt{3\pi}\,\Gamma\left(\textstyle{\frac{11}{6}}\right)/
\Gamma\left(\textstyle{\frac{1}{3}}\right)\big]^{2/5}n^{2/5}\quad(n\to\infty).
\label{E11}
\end{equation}
\end{enumerate}

These asymptotic results for Painlev\'e I have recently been confirmed at a
rigorous level \cite{r5}. However, these behaviors can be easily understood at
a
heuristic level because for large eigenvalues one can approximate the nonlinear
eigenvalue problem associated with the Painlev\'e I equation by the {\it linear}
eigenvalue problem associated with the $\cPT$-symmetric cubic Hamiltonian $H=p^2+
ix^3$ \cite{r4}. To demonstrate this we multiply the Painlev\'e I equation in
(\ref{E9}) by $y'(x)$ and integrate from $0$ to $t$. The result is
$$-\half\left[y'(0)\right]^2+2y^3(0)=-\half\left[y'(t)\right]^2+2y^3(t)+I(t),$$
where $I(t)$ is the indefinite integral
\begin{equation}
I(t)\equiv\int_0^t ds\,sy'(s).
\label{E13}
\end{equation}

\begin{figure}[h!]
\begin{center}
\includegraphics[width=1.0\columnwidth]{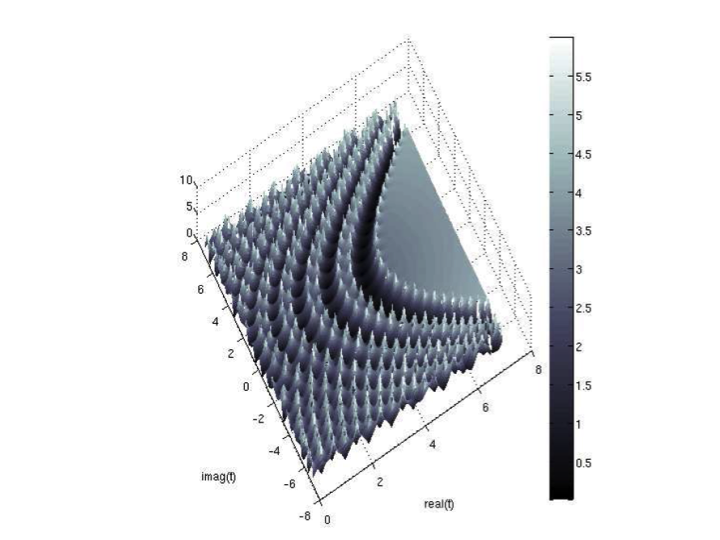}
\caption{\label{F3} A plot of $I(t)$ in (\ref{E13}) showing that for large
eigenvalues this function vanishes as $|t|\to\infty$ along the lines ${\rm arg}
\,t=\pm\quarter\pi$.}
\end{center}
\end{figure}

Figure \ref{F3} shows that as $|t|\to\infty$ in the complex-$t$ plane at an
angle of $\pm\quarter\pi$, the function $I(t)$ vanishes rapidly for large
eigenvalues $E_n$ and thus the Hamiltonian-like quantity $\half\big[y'(t)\big]^2
-2y^3(t)$ is a constant independent of $t$. Hence, we obtain a
quantum-mechanical Hamiltonian whose large eigenvalues are related to the large
eigenvalues of the nonlinear eigenvalue problem for the Painlev\'e equation
\cite{r2}. Since we can use WKB theory to find the large-eigenvalue behavior of
the cubic $\cPT$-symmetric Hamiltonian $H=p^2+ix^3$, we can obtain directly the
results in (\ref{E10}) and (\ref{E11}).

\subsection{Eigenvalue problems for Painlev\'e II}\label{ss1c}
The initial-value problem for the second Painlev\'e equation (Painlev\'e II) has
the form
\begin{equation}
y''(x)=2y^3+xy+Q_0,\quad y(0)~{\rm and}~y'(0)~{\rm specified},
\label{E14}
\end{equation}
where $Q_0$ is a parameter. For simplicity, in our study of this equation we
have taken $Q_0=0$, so for negative $x$ there are three solutions to
$F(x,y)=2y^3+xy=0$:
$$f_0=0\quad{\rm and}\quad f_\pm(x)=\pm\sqrt{-x/2}.$$

The solutions to (\ref{E14}) with $Q_0=0$ and with specified initial conditions
at $x=0$ exhibit three possible behaviors for negative $x$:
\begin{enumerate}
\item After passing through a finite number of simple poles, the solution may
oscillate about the negative-$x$ axis with decreasing
amplitude as $x\to-\infty$ [see Figs.~8 and 9 (right panels) in Ref.~\cite{r2}].
Thus, $f_0(x)$ is a stable fixed point in function space.
\item The solution may have an infinite sequence of simple poles along the
negative-real axis [see Figs.~8 and 9 (left panels) in Ref.~\cite{r2}].
\item After passing through a finite number of simple poles on the negative-$x$
axis, the solution may approach either the upper half or the lower half of the
parabola $2y^2+x=0$ asymptotically [see Figs.~10-12 in Ref.~\cite{r2}]:
$$y(x)\sim\pm\sqrt{-x/2}\quad(x\to-\infty).$$
These solutions are unstable separatrices, so we regard them as eigenfunctions.
\end{enumerate}

There are also interesting separatrix solutions for positive $x$: After passing
through a finite number of poles on the {\it positive}-real axis, the solution
may approach zero monotonically as $x\to\infty$ [see Figs.~13-15 in
Ref.~\cite{r2}]. These solutions are also unstable separatrices, and we again
regard them as eigenfunctions.

The eigenvalue problems for Painlev\'e II have a slightly richer class of
solutions than those for Painlev\'e I.
\begin{enumerate}
\item $y(0)=0$: In this case the values of $y'(0)$ that give rise to
separatrices for $x<0$ are the eigenvalues. For $Q_0=0$ it has been shown [see
Ref.~\cite{r2}] that the large-eigenvalue behavior is
\begin{equation}
y_n'(0)\sim\big[3\sqrt{2\pi}\,\Gamma\left(\threequarter\right)/\Gamma\left(
\quarter\right)\big]^{2/3}n^{2/3}\quad(n\to\infty).
\label{E15}
\end{equation}
This asymptotic behavior can be obtained by solving the linear eigenvalue
problem associated with the quartic $\cPT$-symmetric Hamiltonian $H=p^2-x^4$
\cite{r4}.
\item $y'(0)=0$: In this case the values of $y(0)$ that give rise to
separatrices for $x>0$ are the eigenvalues. Now, the large-eigenvalue behavior
for $Q_0=0$ is
\begin{equation}
y_n(0)\sim\big[3\sqrt{\pi}\,\Gamma\left(\threequarter\right)/\Gamma\left(
\quarter\right)\big]^{1/3}n^{1/3}\quad(n\to\infty).
\label{E16}
\end{equation}
This asymptotic behavior can be obtained by solving the linear eigenvalue
problem associated with the Hermitian quartic anharmonic oscillator Hamiltonian
$H=p^2+x^4$.
\end{enumerate}

\subsection{Summary of results in this paper}\label{ss1d}
The objective of this paper is to extend and generalize the asymptotic results
above. The simplest extension concerns the higher-order corrections in powers
of $1/n$ to the leading-order asymptotic approximations above for Painlev\'e I
and II. Second, we examine the effect of having inhomogeneous initial
conditions. This work is presented in Sec.~\ref{s2}.

One generalization involves constructing and analyzing new kinds of nonlinear
differential equations. The Painlev\'e equations are special because the
solutions are {\it meromorphic}; that is, their movable singularities are poles 
so they live on just one sheet of a Riemann surface. Almost always, if one
generalizes the Painlev\'e equations [for example, if one were to replace
Painlev\'e I in (\ref{E9}) by $y''(x)=6y^2(x)+e^x$], the movable singularities
have logarithmic structure so the solutions now live on an infinite-sheeted
Riemann surface \cite{r6}. A well known nonlinear equation whose solutions live
on an infinite-sheeted Riemann surface is the Thomas-Fermi equation $y''(x)=
y^{3/2}(x)/\sqrt{x}$. The movable singularities of this equation may seem to be
fourth-order poles, but when we attempt to construct a Laurent series near a
movable singularity, we discover a complicated logarithmic structure that first
appears in {\it tenth} order. A detailed discussion of this equation is given
in Sec.~\ref{s3}.

In Sec.~\ref{s4} we identify a large class of nonlinear differential equations
whose movable singularities are {\it algebraic} rather than logarithmic in
character:
\begin{equation}
y''(x)=Ay^M(x)+xp(y)+q(y),
\label{E17}
\end{equation}
where $M\geq2$ is an integer, $A$ is a numerical constant, and $p(y)$ and $q(y)$
are polynomials in $y$ of degree at most $M-2$. Painlev\'e I and II are special
cases of this class of equations with $M=2$ and $M=3$, and for these values of
$M$ the movable singularities are poles. When $M\geq4$, the movable
singularities are algebraic branch points at which a {\it finite} number of
Riemann sheets are joined together.

In general, the solutions to (\ref{E17}) have an infinite number of
singularities on the negative-real axis. However, in Sec.~\ref{s5} we present an
extensive numerical study that shows that there are an infinite number of
discrete initial conditions (eigenvalues) for which the solutions on the
negative-real axis remain real and only have a finite number of singularities.
These eigenvalues have the algebraic asymptotic form $E_n\sim\alpha n^\beta$ for
large $n$. For this general class of nonlinear equations we determine the values
of $\beta$ as simple rational numbers. This is a principal result of this paper.
So far, closed-form analytic expressions for $\alpha$ are known only for
Painlev\'e I and II; for $M\geq4$ only numerical results for $\alpha$ are
available.

Some of the generalized Painlev\'e equations in (\ref{E17}) possess new
features that Painlev\'e I and II do not display. For example, the eigenvalues
may exhibit a hyperfine structure like that seen in atomic physics. In
Sec.~\ref{s6} we describe this hyperfine structure in detail and present a
heuristic asymptotic analysis for the size of the splitting. 

The first six sections of this paper focus on eigenvalue data for large classes
of nonlinear eigenvalue problems, and most of this data is numerical. Section
\ref{s7} concludes by proposing a range of avenues for future research.

\section{Higher approximations to eigenvalues of Painlev\'e I and II}
\label{s2}
It is natural to investigate the higher-order corrections to the asymptotic
approximations of the form $E_n\sim\alpha n^\beta$ in (\ref{E10}--\ref{E11}) and
(\ref{E15}--\ref{E16}). This section presents some new numerical results
regarding the form of the asymptotic behavior of the eigenvalues $E_n$ for large
$n$. We find that, in general, the large-$n$ behavior of $E_n$ takes the form
$$E_n\sim\alpha(n-\gamma)^\beta\left[1+\frac{\delta_2}{(n-\gamma)^2}+\frac{
\delta_3}{(n-\gamma)^3}+\frac{\delta_4}{(n-\gamma)^4}+\frac{\delta_5}{(n-\gamma
)^5}+\frac{\delta_6}{(n-\gamma)^6}+\cdots\right],$$
where $\gamma$ and $\delta_k$ are constants. Note that the term $\delta_1/(n-
\gamma)$ does not appear in this series.

\subsection{Painlev\'e I: $y''(x)=6y^2(x)+x$}\label{ss2a}
For the initial-slope problem with $y(0)=0$, we find that for large $n$ the
behavior of the positive eigenvalues is given by
$$y_n'(0)\sim\alpha(n-\gamma)^{3/5}\left[\textstyle{1-\frac{0.005\,513\,28}{(n-
\gamma)^2}+\frac{0.29\,334}{(n-\gamma)^3}+\frac{0.0359}{(n-\gamma)^4}+
\frac{4.54}{(n-\gamma)^5}-\frac{0.38}{(n-\gamma)^6}}\right],$$
where $\alpha=2\left[\sqrt{3\pi}\Gamma\left(\textstyle{\frac{11}{6}}\right)/
\Gamma\left(\third\right)\right]^{3/5}\approx 2.092\,146\,744\,884\,417$ and
$\gamma=\frac{1}{6}$ is a simple rational number; this is because we have taken
$y(0)=0$. The large-$n$ behavior of the negative eigenvalues is given by
$$y_n'(0)\sim-\alpha(n-\gamma)^{3/5}\left[\textstyle{1-\frac{0.005\,52}{(n-
\gamma)^2}-\frac{0.3}{(n-\gamma)^3}}\right],$$
where in this case $\gamma=\frac{5}{6}$. 

For an inhomogeneous initial condition $y(0)$, the parameter $\gamma$ is no
longer simple. For example, with the inhomogeneous initial condition $y(0)=1$,
the positive eigenvalues $y_n'(0)$ are approximated by the expansion
$$\half[y'_n(0)]^2-2\sim\alpha(n-\gamma)^{6/5}\left[\textstyle{1+\frac{0.33121}
{(n-\gamma)^2}-\frac{10.40}{(n-\gamma)^3}}\right]$$
with $\alpha\approx2.188356$ and $\gamma\approx1.147996$.
(Note that the left side of this expression has a complicated form. The origin
of this structure is explained in Sec.~\ref{s5}.) Observe that $\alpha$ here is
very slightly different from the value of $\alpha=2\big[\sqrt{\pi/3}\Gamma\left(
\textstyle{\frac{11}{6}}\right)/\Gamma\left(\third\right)\big]^{6/5}\approx
2.188\,539\,001\,065\,231$ in the problem with a homogeneous initial condition.
However, $\gamma$ here is quite different from the previous value $1/6$.

For the inhomogeneous initial condition $y(0)=-1$ the positive eigenvalues
$y'_n(0)$ are approximated by the expansion
$$\half[y'_n(0)]^2+2\sim\alpha(n-\gamma)^{6/5}\left[\textstyle{1-\frac{0.712}
{(n-\gamma)^2}-\frac{5.28}{(n-\gamma)^3}}\right]$$
with $\alpha\approx2.18878$ and $\gamma\approx-0.8115$. Note that in this case
$\gamma$ becomes negative.

For the alternative eigenvalue problem with $y'(0)=0$ the eigenvalues [the
values of $y(0)$] are only negative. For large $n$ we find that
$$y_n(0)\sim-\alpha(n-\gamma)^{2/5}\left[\textstyle{1-\frac{0.009\,651\,8}{(n
-\gamma)^2}+\frac{0.0240}{(n-\gamma)^4}}\right].$$
Here, $\alpha=\left[\sqrt{3\pi}{\Gamma\left(\frac{11}{6}\right)}/\Gamma\left(
\third\right)\right]^{2/5}\approx1.030\,484\,423\,696\,866$ and $\gamma=\half$.
We believe that there are no odd powers of $(n-\gamma)$ in this asymptotic
approximation. All three values of $\gamma$ above with homogeneous initial 
conditions are in agreement with the findings in Ref.~\cite{r5}.

For the inhomogeneous initial condition $y'(0)=1$ the eigenvalues $y_n(0)$ are
approximated by the expansion
$$\half-2[y_n(0)]^3\sim2.1885n^{6/5}\left[\textstyle{1-\frac{0.59465}{n}-\frac{
0.676}{n^2}-\frac{7.95}{n^3}}\right].$$
Also, for the inhomogeneous initial condition $y'(0)=2$ the eigenvalues $y_n(0)$
are approximated by
$$2-2[y_n(0)]^3\sim2.1885n^{6/5}\left[\textstyle{1+\frac{0.608}{n}+\frac{1.64}
{n^2}-\frac{15.6}{n^3}}\right].$$
In these last two equations we have not bothered to introduce the shift parameter
$\gamma$.

\subsection{Painlev\'e II: $y''(x)=2y^3(x)+xy(x)$}\label{ss2b}
For the initial-slope problem with $y(0)=0$ we find that for large $n$ the
behavior of the odd eigenvalues is given by
$$y_{2n+1}'(0)\sim\alpha(n-\gamma)^{2/3}\left[\textstyle{1-\frac{0.004\,
420\,9}{(n-\gamma)^2}-\frac{0.041}{(n-\gamma)^4}}\right],$$
where $\alpha=\left[\sqrt{\textstyle{\frac{\pi}{2}}}{\Gamma\big(\textstyle{\frac
{7}{4}}\big)}/{\Gamma\big(\textstyle{\frac{5}{4}}\big)}\right]^{2/3}\approx
1.173\,246\,522\,889\,079$ and $\gamma=\threequarter$. The even eigenvalues have
a similar expansion:
$$y_{2n}'(0)\sim\alpha(n-\gamma)^{2/3}\left[\textstyle{1-\frac{0.004\,
420\,9}{(n-\gamma)^2}-\frac{0.0415}{(n-\gamma)^4}}\right],$$
where in this case $\gamma=\quarter$. In both equations above the odd powers of
$n-\gamma$ appear to vanish to all orders.

For an inhomogeneous initial condition $y(0)\neq0$, again $\gamma$ is not
simple. Thus, if $y(0)=1$, the eigenvalues $y_{2n+1}'(0)$ are approximated by
the expansion
$$\half[y'_{2n+1}(0)]^2-\half\sim\alpha(n-\gamma)^{4/3}\left[\textstyle{1+
\frac{0.575}{(n-\gamma)^2}-\frac{12.7}{(n-\gamma)^3}}\right]$$
with $\alpha\approx1.73415$ and $\gamma\approx1.2281$. Note that this value of 
$\alpha$ is slightly different from $\alpha=\half\left[\sqrt{2\pi}\Gamma\left(
\textstyle{\frac{7}{4}}\right)/\Gamma\left(\third\right)\right]^{4/3}\approx
1.734\,290\,652\,969\,643$, which is associated with the homogeneous initial
condition. Note also that the value of $\gamma$ for the initial condition $y(0)=
1$ is quite different from the value $3/4$ associated with the initial condition
$y(0)=0$. The even-numbered eigenvalues are approximated by
$$\half[y'_{2n}(0)]^2-\half\sim\alpha(n-\gamma)^{4/3}\left[\textstyle{1+
\frac{0.588}{(n-\gamma)^2}-\frac{13.5}{(n-\gamma)^3}}\right]$$
with $\alpha\approx1.73415$ and $\gamma\approx0.7282$.

For the different initial condition $y(0)=-1$ the odd and even eigenvalues $y'_n
(0)$ for large $n$ are given approximately by the expansions
$$\half[y'_{2n+1}(0)]^2-\half\sim\alpha(n-\gamma)^{4/3}\left[\textstyle{1-
\frac{0.65}{(n-\gamma)^2}-\frac{14.2}{(n-\gamma)^3}}\right]$$
with $\alpha\approx1.73441$ and $\gamma\approx0.271$ and 
$$\half[y'_{2n}(0)]^2-\half\sim\alpha(n-\gamma)^{4/3}\left[\textstyle{1-
\frac{0.520}{(n-\gamma)^2}-\frac{9.97}{(n-\gamma)^3}}\right]$$
with $\alpha\approx1.734455$ and $\gamma\approx-0.227$.

For the alternative eigenvalue problem with $y'(0)=0$ we find that
$$y_n(0)\sim\alpha n^{1/3}\left(\textstyle{1+\frac{0.005\,254\,3}{n^2}+
\frac{0.077}{n^4}}\right).$$
Here, $\alpha=\left[\sqrt{\pi}\Gamma\left(\textstyle{\frac{7}{4}}\right)/\Gamma
\left(\textstyle{\frac{5}{4}}\right)\right]^{1/3}\approx1.215\,811\,659\,305\,
798$ and $\gamma=0$.

Again, for the inhomogeneous initial condition $y'(0)=1$, the eigenvalues
$y_n(0)$ are approximated by the expansion
$$\half-\half[y_n(0)]^4\sim-1.09254n^{4/3}\left[\textstyle{1+\frac{0.57}{n^2}
-\frac{28}{n^3}}\right].$$
Note that the prefactor here is slightly different from $\half\left[\sqrt{\pi}
\Gamma \left(\textstyle{\frac{7}{4}}\right)/\Gamma\left(\third\right)\right]^{
4/3}\approx 1.092\,534\,650\,156\,189$, which is the prefactor associated
with the homogeneous initial condition. For the inhomogeneous initial condition
$y'(0)=2$ the eigenvalues $y_n(0)$ are given by the expansion
$$2-\half[y_n(0)]^4\sim-1.09255n^{4/3}\left[\textstyle{1+\frac{1.05}{n^2}
-\frac{24.5}{n^3}}\right].$$
Note that in these cases the value of $\gamma$ for the inhomogeneous initial
condition is unchanged from that for the homogeneous case.

\section{Difficulties with the Thomas-Fermi equation}\label{s3}
The Thomas-Fermi equation gives a semiclassical description of the charge
density in a nucleus. This equation is posed as the two-point boundary-value
problem
\begin{equation}
y''(x)=y^{3/2}(x)/\sqrt{x},\quad y(0)=1,~y(+\infty)=0.
\label{E24}
\end{equation}
There is a unique solution to this boundary-value problem that is positive for
all $x\geq0$ and decays monotonically to $0$ as $x$ increases. The leading
asymptotic behavior of $y(x)$ is
$$y(x)\sim144x^{-3}\quad(x\to+\infty).$$

If the initial slope of $y(x)$ is $b$, $y'(0)=b$, then the small-$x$
behavior of $y(x)$ is given by
\begin{eqnarray}
y(x)&\sim&1+bx+\frac{4}{3}x^{3/2}+\frac{2b}{5}x^{5/2}+\frac{1}{3}x^3+\frac{3b^2}
{70}x^{7/2}+\frac{2b}{15}x^4+\left(\frac{2}{27}-\frac{b^3}{252}\right)x^{9/2}
+\frac{b^2}{175}x^5\nonumber\\
&& \quad+b\left(\frac{31}{1485}+\frac{b^3}{1056}\right)x^{11/2}+\left(\frac{4}
{405}+\frac{4b^3}{1575}\right)x^6+b^2\left(\frac{557}{100\,100}-\frac{3b^3}
{9152}\right)x^{13/2}\nonumber\\
&&\quad+b\left(\frac{4}{693}-\frac{9b^3}{24\,255}\right)x^7+\left(\frac{101}
{52\,650}-\frac{623b^3}{351\,000}+\frac{7b^6}{49\,920}\right)x^{15/2}\nonumber\\
&&\quad -b^2\left(\frac{46}{45\,045} -\frac{68b^3}{105\,105}\right)x^8+\cdots.
\nonumber
\end{eqnarray}
The objective of the Thomas-Fermi boundary-value problem is to determine the
value of $b$.

The unique solution to the boundary-value problem (\ref{E24}) is an
eigenfunction in the same sense as the Painlev\'e eigenfunctions discussed in
Sec.~\ref{s2}. For the Thomas-Fermi equation the eigenvalue is the initial slope
$y'(0)=b$. The unique solution is a separatrix that is unstable relative to
small
changes in $b$. If $b$ is increased by a small amount, the solution blows up at
some finite positive value $x=a$; the eigenfunction solution is the lower
bound of such solutions that blow up. On the other hand, if $b$ is decreased
by a small amount, the solution passes through 0 and becomes complex; the
eigenfunction is the upper bound of all such solutions.

If the solution to the Thomas-Fermi equation becomes infinite at $x=a$, the
leading asymptotic behavior of $y(x)$ in the neighborhood of this movable
singularity is given by
\begin{equation}
y(x)\sim400a(x-a)^{-4}\quad(x\to a).
\label{E26}
\end{equation}
Thus, while the solution approaches $+\infty$ as $x$ approaches $a$ from below,
it appears to come back down to finite positive values when $x>a$. Based
on the behavior of the solutions to the Painlev\'e transcendents (see
Ref.~\cite{r2}), one might think that there are more separatrix solutions that
are generated by larger values of $b$. That is, one might guess that there is a
sequence of initial slopes $b_0,\,b_1,\,b_2,\,b_3,\,\ldots$ for which there are
separatrix solutions $y_n(x)$ that (i) satisfy $y_n(0)=1$ and $y_n'(0)=b_n$,
(ii) pass through $n$ movable singularities of the form in (\ref{E26}), (iii)
remain positive for all $x$, and (iv) approach $0$ as $x\to\infty$.

However, the situation for the Thomas-Fermi equation is more complicated than
that for the Painlev\'e transcendents. Let us examine the higher-order
corrections to the asymptotic behavior near a movable singularity at $x=a$. If
we seek an expansion in the form of a Laurent series
$$y(x)=\frac{400a}{(x-a)^4}\,\sum_{n=0}^\infty C_n(x-a)^n,$$
we find that this expansion is insufficient unless we include in addition a
logarithm term at tenth order. The correct expansion to order eleven is
\begin{eqnarray}
y(x) &\sim& \frac{400\,a}{(x-a)^4}\left\{1+\frac{5(x-a)}{9a}-\frac{5(x-a)^2}{81
a^2}+\frac{25(x-a)^3}{729a^3}-\frac{625(x-a)^4}{26\,244a^4}\right.\nonumber\\
&&+\frac{4447(x-a)^5}{236\,196a^5}-\frac{81\,275(x-a)^6}{4\,960\,116a^6}
+\frac{696\,575(x-a)^7}{44\,641\,044a^7}-\frac{80\,213\,375(x-a)^8}
{4\,821\,232\,752a^8}\nonumber\\
&&+\frac{137\,618\,915(x-a)^9}{137\,618\,915a^9}+\left[C_{10}+\log(x-a)
\right]\frac{44\,232\,230(x-a)^{10}}{4\,261\,625\,379a^{10}}\nonumber\\
&&\left.-\left(\frac{6\,331\,279\,880}{805\,447\,196\,631}
+\frac{221\,161\,150}{38\,354\,628\,411}\left[C_{10}+\log(x-a)\right]\right)
\frac{(x-a)^{11}}{a^{11}}+\cdots\right\}.
\label{E27}
\end{eqnarray}
Furthermore, one additional power of $\log(x-a)$ appears every ten terms in the
series. Thus, the movable singularity is not a fourth-order pole but rather a
complicated logarithmic branch point.

When the logarithm term first appears in tenth order, it is accompanied by the
arbitrary coefficient $C_{10}$. The appearance of an arbitrary coefficient in
this series expansion is necessary and expected because the order of the
differential equation is two, and the general solution must contain two
arbitrary constants. The second arbitrary constant is $a$, the location of the
movable singularity. By comparison, for the Painlev\'e I equation the expansion
around a movable singularity at $x=a$ has the form
$$y(x)=\frac{1}{(x-a)^2}\,\sum_{n=0}^\infty C_n(x-a)^n,$$
where all of the coefficients are uniquely determined except for $C_6$, which is
arbitrary. The Painlev\'e equation is special because no logarithm term appears
along with the coefficient $C_6$. Also, the series has a nonzero radius of
convergence, which implies that the movable singularity is a double pole.

We conclude that if there are separatrix (eigenfunction) solutions to the
Thomas-Fermi equation, these solutions cease to be real after they pass a
movable singularity and they do not live on just one sheet of a Riemann surface.
It would be extremely interesting if one could find a class of complex
separatrix solutions with different eigenvalues $b_n$, but this would require
much further analysis.

\section{Generalized Painlev\'e equations}\label{s4}
The analysis of the Thomas-Fermi equation in Sec.~\ref{s3} teaches us that it is
dangerous to seek separatrix (eigenfunction) solutions to nonlinear differential
equations beyond the Painlev\'e transcendents. This is because the movable
singularities will not be poles. These singularities are usually complicated
logarithmic branch points, so finding a real separatrix solution would be
difficult or perhaps even impossible. Nevertheless, in this section we show that
there exists an infinite class of nonlinear differential equations whose movable
singularities are algebraic and not logarithmic branch points. At these movable
singularities only a finite number rather than an infinite number of Riemann
sheets are joined. We call this class of equations {\it generalized Painlev\'e
equations}.

The generalized Painlev\'e equations are labeled by an integer $M\geq2$ and have
the form
\begin{equation}
y''(x)=\textstyle{\frac{2(M+1)}{(M-1)^2}}\left[y^M(x)+xp(y)+q(y)\right],
\label{E29}
\end{equation}
where $p(y)$ and $q(y)$ are polynomials in $y$ of degree at most $M-2$:
\begin{eqnarray}
p(y)&=& P_{M-2}\,y^{M-2}(x)+P_{M-3}\,y^{M-3}(x)+\cdots+P_1\,y(x)+P_0,\nonumber\\
q(y)&=& Q_{M-2}\,y^{M-2}(x)+Q_{M-3}\,y^{M-3}(x)+\cdots+Q_1\,y(x)+Q_0.\nonumber
\end{eqnarray}
This class of equations contains Painlev\'e I ($M=2$) and Painlev\'e II ($M=3$)
as special cases.

For even $M=2K$, $p(y)$ and $q(y)$ are arbitrary. The asymptotic behavior of
$y(x)$ near a movable singularity $a$ is given by
$$y(x)\sim\frac{1}{(x-a)^{2/(M-1)}}\left[1+\sum_{n=1}^{\infty}C_n
(x-a)^{n/(M-1)}\right]\quad(x\to a).$$
All of the coefficients $C_n$ are determined by the nonlinear differential
equation (\ref{E29}) except for $C_{2(M+1)}$, which is in principle determined
by the initial conditions. [The coefficient $C_{2(M+1)}$ is analogous to the
coefficient $C_{10}$ in (\ref{E27}) for the Thomas-Fermi equation.] Note that
because $M$ is even, all of the fractional powers in this series can give rise
to real numbers. Thus, {\it there exists a real solution on both sides of the
movable singularity.}

For odd $M=2K+1$, the leading asymptotic behavior of $y(x)$ near a movable
singularity at $x=a$ is $y(x)\sim\pm(x-a)^{-1/K}$. For the positive case the
full asymptotic series is
$$y(x)\sim\frac{1}{(x-a)^{1/K}}\left[1+\sum_{n=1}^{\infty}C_n(x-a)^{n/K}\right].
$$
In this case $C_{2(K+1)}=C_{M+1}$ is the one coefficient that is determined by
the initial conditions. (If there is a negative sign in the leading term, the
resulting series has a similar structure.) Note that a real solution is possible
only if $K$ is odd; when $K$ is even, the solution inevitably becomes complex
when it passes through a movable singularity. Furthermore, to have real
solutions the coefficients in the polynomial functions $p(y)$ and $q(y)$ must
satisfy some constraints, as we illustrate below for some small values of $M$.

For $M=3$ it is necessary that $P_0=0$. One may then scale $x$ to make $P_1=
\half$ and one may also shift $x$ to make $Q_1=0$. Thus, the most general form
for the $M=3$ equation has only one arbitrary constant: $y''(x)=2y^3(x)+xy(x)+
Q_0$. This is the standard form of the Painlev\'e II equation (\ref{E14}).

For $M=5$ we must have $P_1=P_3=0$. This leads to two standard forms for the
generalized Painlev\'e equation for $M=5$:
$$y''(x)=\threequarter\left[y^5(x)+x+Q_3\,y^3(x)+Q_2\,y^2(x)+Q_1\,y(x)\right]$$
and
$$y''(x)=\threequarter\left[y^5(x)+xy^2(x)+Q_3\,y^3(x)+Q_1\,y(x)+Q_0\right].$$
For higher values of $M$ the results are similar. For $M=7$ either
$$P_5=0,~~P_2=\textstyle{\frac{2}{5}}P_4 Q_5\quad{\rm or}\quad
P_4=0,~~P_2=\textstyle{\frac{2}{5}}P_5 Q_4,$$
and for $M=9$, 
$$P_6=P_7=0,\quad P_3=\textstyle{\frac{5}{12}}P_5 Q_7.$$
There are three choices for $M=11$, and so on.

To illustrate the special properties of the generalized Painlev\'e equations
we consider the simple case for which $M=4$: $p(y)=y^2$, and $q(y)=0$:
\begin{equation}
y''(x)=\textstyle{\frac{10}{9}}\left[y^4(x)+xy^2(x)\right].
\label{E32}
\end{equation} 
First, we observe that the solutions to (\ref{E32}) are regular near the origin.
For the initial conditions $y(0)=C_0$ and $y'(0)=C_1$ the Taylor expansion about
$x=0$ is
\begin{eqnarray}
y(x)&=&\textstyle{C_0+C_1x+\frac{5}{9}C_0^4x^2+\left(\frac{5}{27}C_0^2+
\frac{20}{27}C_0^3C_1\right)x^3+\left(\frac{50}{243}C_0^7+\frac{5}{27}C_0C_1+
\frac{5}{9}C_0^2C_1^2\right)x^4}\nonumber\\
&&\quad\textstyle{+\left(\frac{25}{243}C_0^5+\frac{130}{243}C_0^6C_1+\frac{1}
{18}C_1^2+\frac{2}{9}C_0 C_1^3\right)x^5+\cdots}.\nonumber
\end{eqnarray}
The initial conditions play the role of eigenvalues. As in the case of the
Painlev\'e equations, there is one eigenvalue problem for which $C_0$ is held
fixed and $C_1$ is the eigenvalue that yields a separatrix solution. There is a
second eigenvalue problem for which $C_1$ is held fixed and $C_0$ is the
eigenvalue that yields a separatrix solution.

Second, for large negative $x$ the leading asymptotic behavior of the separatrix
solution is $y(x)\sim\pm\sqrt{-x}$. For the positive sign the full asymptotic
series has the general form
$$y(x)\sim\sqrt{-x}\left[1+\sum_{n=1}^\infty C_n\left(-\frac{1}{x}\right)^{7n/2}
\right]\quad (x\to-\infty)$$
and up to the sixth order this expansion for large negative $x$ reads
\begin{eqnarray}
y(x) &\sim& \sqrt{-x}\left[1-\frac{9}{80}\left(-\frac{1}{x}\right)^{7/2}-\frac{
8181}{12\,800}\left(-\frac{1}{x}\right)^7-\frac{47\,235\,873\,573}{65\,536\,000}
\left(-\frac{1}{x}\right)^{14}\right.\nonumber\\
&&\quad\left.-\frac{209\,519\,094\,269\,691}{3\,276\,800\,000}\left(-\frac{1}{x}
\right)^{35/2}-\frac{37\,091\,396\,224\,409\,411\,997}{4\,194\,304\,000\,000}
\left(-\frac{1}{x}\right)^{21}\right]+\cdots.\nonumber
\end{eqnarray}

Third, near a movable singularity at $x=a$, the solution blows up algebraically
and has the general form
$$y(x)=\frac{1}{(x-a)^{2/3}}\left[1+\sum_{n=1}^\infty C_n(x-a)^{n/3}\right].$$
Near $x=a$ the first 21 terms in this series are
\begin{eqnarray}
y(x) &\sim& \frac{1}{(x-a)^{2/3}}\left[1-\frac{5a}{21}(x-a)^{4/3}-\frac{1}{3}(
x-a)^{7/3}+\frac{100a^2}{1617}(x-a)^{8/3}+C_{10}(x-a)^{10/3}\right.\nonumber\\
&&\quad-\frac{20a}{147}(x-a)^{11/3}-\frac{5125a^3}{305\,613}(x-a)^4-\frac{
15aC_{10}}{119}(x-a)^{14/3}-\frac{625a^2}{305\,613}(x-a)^5\nonumber\\
&&\quad+\frac{6\,254\,375 a^4}{1\,341\,335\,457}(x-a)^{16/3}-\frac{C_{10}}{7}
(x-a)^{17/3}+\left(\frac{55a}{4116}+\frac{575a^2C_{10}}{9163}\right)(x-a)^6
\nonumber\\
&&\quad-\frac{101\,000a^3}{70\,596\,603}(x-a)^{19/3}+\left(\frac{
13\,550\,000a^5}{10\,283\,571\,837}-\frac{6C_{10}^2}{23}\right)(x-a)^{20/3}
\nonumber\\
&&\quad\left.-\left(\frac{5}{3564}+\frac{1550aC_{10}}{82\,467}\right)(x-a)^7
+\cdots\right].\nonumber
\end{eqnarray}
Observe that the new arbitrary parameter $C_{10}$ is {\it not accompanied by a
logarithmic term}. Thus, the generalized Painlev\'e equation (\ref{E32}) evades
the problem presented by the Thomas-Fermi equation, where a logarithmic term
appears in the series (\ref{E27}).

\section{Eigenvalue problems for generalized Painlev\'e equations}\label{s5}
In this section we examine the eigenvalues for the special case of the
generalized Painlev\'e equation with just three terms,  
\begin{equation}
y''(x)=Ay^M(x)+Bxy^{M-m}(x),
\label{E35}
\end{equation}
where $A$ and $B$ are constants and $M>1$ and $m>1$ are integers. For each
differential equation, we consider two types of nonlinear eigenvalue problems.
For the first, we fix the initial value $y(0)=0$ and treat the initial slope as
the eigenvalue when a separatrix solution arises. We call this the {\it
initial-slope eigenvalue problem.} For the second, we fix the initial slope
$y'(0)=0$ and treat the initial value $y(0)$ as the eigenvalue. We call this the
{\it initial-function eigenvalue problem.} For some generalized Painlev\'e
equations the separatrix solutions as $x\to+\infty$ and as $x\to-\infty$ are
different.
 
\subsection{Relation between initial-slope problems and initial-function
problems}\label{ss5a}
To establish a relation between these two types of eigenvalue
problems, we multiply (\ref{E35}) by $y'(x)$ and integrate:
\begin{equation}
-\half\left[y'(0)\right]^2+\frac{A}{M+1}y^{M+1}(0)=-\half\left[y'(x)\right]^2+
\frac{A}{M+1}y^{M+1}(x)+B\int_0^x ds\,sy^{M-m}(s)y'(s).
\label{E36}
\end{equation}
For the equations considered here, the right side of (\ref{E36}) becomes
independent of $x$ to leading order in the large-$n$ expansion. Depending on the
oddness of $M$ and the sign of the power-law behavior $\alpha n^\beta$,
different eigenvalue problems may become related. 

\subsubsection{Even $M$}
\begin{itemize}
\item {Case A: Positive right side.} Suppose that for large $n$
$$\half\left[y_n'(0)\right]^2-\frac{A}{M+1}y_n^{M+1}(0)\sim\alpha n^\beta\quad
\quad(n\to+\infty).$$
This leads to
\begin{eqnarray}
y_n'(0) &\sim& \pm\sqrt{2\alpha n^\beta+\textstyle{\frac{2A}{M+1}}y^{M+1}(0)},
\nonumber\\
y_n(0) &\sim& -\left\{\textstyle{\frac{\alpha(M+1)}{A}}n^\beta-\textstyle{\frac
{M+1}{2A}}\left[y'(0)\right]^2\right\}^{1/(M+1)}.
\label{E38}
\end{eqnarray}
For example, to leading order the eigenvalues of Painlev\'e I  satisfy
\begin{equation}
-\half\left[y_n'(0)\right]^2+2y_n^3(0)\sim-\alpha n^\beta,
\label{E39}
\end{equation}
where $\alpha=2\big[\sqrt{\pi/3}\Gamma\left(\textstyle{\frac{11}{6}}\right)/
\Gamma\left(\third\right)\big]^{6/5}\approx2.188\,539\,001\,065\,231$ and $\beta
=6/5$. Thus, the eigenvalues of the initial-slope problem satisfy
$$y_n'(0)\sim\pm\sqrt{2\alpha n^\beta+4y^3(0)}$$ 
for a fixed $y(0)$. Similarly, the eigenvalues for the initial-function problem
satisfy
$$y_n(0)\sim-\left\{\half\alpha n^\beta-\quarter\left[y'(0)\right]^2\right\}^{
1/3}$$
with fixed $y'(0)$. For example, for the eigenvalue problems of Painlev\'e I
with a homogeneous boundary condition, we find that to leading order, $\half
\big[y_n'(0)\big]^2\sim2.188\,538\,91\,n^{6/5}$ and $-2y_n^3(0)\sim2.188\,538\,
85\,n^{6/5}$. This is in good agreement with (\ref{E39}).

\item {Case B: Negative right side.} Suppose the right side has a negative
asymptotic behavior:
$$\half\left[y_n'(0)\right]^2-\frac{A}{M+1}y_n^{M+1}(0)\sim-\alpha n^\beta\quad
{\rm as}\quad(n\to+\infty).$$
This leads to
$$y_n(0)\sim\left\{\textstyle{\frac{\alpha(M+1)}{A}}n^\beta+\textstyle{\frac{
M+1}{2A}}\left[y'(0)\right]^2\right\}^{1/(M+1)}.$$
In this case, there are no real eigensolutions for the initial-slope problem.
\end{itemize}

\subsubsection{Odd $M$}
\begin{itemize}
\item {Case A: Positive right side.} Suppose that as $n\to+\infty$,
$\half\left[y_n'(0)\right]^2-\frac{A}{M+1}y_n^{M+1}(0)\sim\alpha n^\beta$.
This leads to
$$y_n'(0)\sim\pm\sqrt{2\alpha n^\beta+\textstyle{\frac{2A}{M+1}}y^{M+1}(0)}.$$
In this case there are no real eigensolutions for the initial-function problem.
Thus, Painlev\'e II for negative $x$ has eigensolutions only for the
initial-slope problem and no eigensolutions for the initial-function problem. 

\item {Case B: Negative right side.} Suppose that
$$\half\left[y'(0)\right]^2-\frac{A}{M+1}y^{M+1}(0)\sim-\alpha n^\beta\quad
(n\to+\infty).$$
This leads to
$$y_n(0)\sim\pm\left\{\textstyle{\frac{\alpha(M+1)}{A}}n^\beta+\textstyle{\frac
{M+1}{2A}}\left[y'(0)\right]^2\right\}^{1/(M+1)}.$$
In this case, there are no eigensolutions for the initial-slope problem. This is
exactly what we find for Painlev\'e II in the positive-$x$ domain. The
initial-function problem has nontrivial eigensolutions but no eigenvalues for
the initial-slope problem. 
\end{itemize}
We emphasize that the simple relation between the two eigenvalue problems only
holds to leading order for large $n$.  

\subsection{Asymptotic behavior of the eigenvalues}
Below we present the numerical results of our extensive study of the nonlinear
eigenvalue problems for various generalized Painlev\'e equations.

\subsubsection{Generalized Painlev\'e 4a (GP4a): $y''(x)=\frac{10}{9}y^4(x)
+xy^2(x)$}
For the initial-slope problem with $y(0)=0$, the large-$n$ behaviors of the
eigenvalues of the separatrix eigensolutions for negative $x$ are found to be
$$y_n'(0)\sim 2.9996(n-0.192)^{5/7},\quad y_n'(0)\sim-2.9996(n-0.604)^{5/7}.$$
As predicted in (\ref{E38}), the initial-function problem with $y'(0)=0$ on the
same side of the $x$-axis has only negative eigenvalues. For large $n$,
$$y_n(0)\sim -1.82502(n-0.42)^{2/7}.$$
To leading order $\half\left[y_n'(0)\right]^2\sim 4.4988\,n^{10/7}$ is close to
$-\frac{2}{9}y_n^5(0)\sim4.4991\,n^{10/7}$. For positive $x$ only the
initial-function problem has nontrivial eigenvalues, as expected. For large $n$,
$$y_n(0)\sim 1.098102(n-1.00104)^{2/7}.$$

\subsubsection{Generalized Painlev\'e 4b (GP4b): $y''(x)=\frac{10}{9}y^4(x)+x
y(x)$}
The eigenvalues for the initial-slope problem with $y(0)=0$ have the large-$n$
behavior
\begin{eqnarray}
y_n'(0) &\sim& 2.1336 \left\{\begin{array}{ll}(n-0.71)^{5/9} & (n\,{\rm odd}),\\
(n-0.43)^{5/9} & (n\,{\rm even}),\end{array}\right.\nonumber\\
y_n'(0) &\sim& -2.1336 \left\{ \begin{array}{ll}(n-0.65)^{5/9} & (n\,{\rm odd}),
\\ (n-0.41)^{5/9} & (n\,{\rm even}).\end{array}\right.\nonumber
\end{eqnarray}
Note that the first-order corrections are different for even $n$ and odd $n$. 

The eigenvalues of the initial-function problem with $y'(0)=0$ have the
asymptotic behavior
\begin{eqnarray}
y_n(0) &\sim& -1.59255\left\{\begin{array}{ll}(n-0.62)^{2/9} & (n\,{\rm odd})\\
(n-0.38)^{2/9} & (n\,{\rm even}).\end{array}\right.\nonumber
\end{eqnarray}
To leading order $\half\left[y_n'(0)\right]^2\sim 2.2761\,n^{10/9}$, which
agrees with $-\frac{2}{9}y_n^5(0)\sim 2.27642\,n^{10/9}$. For the remainder of
this section we only present numerical results.

\subsubsection{Generalized Painlev\'e 4c (GP4c): $y''=\frac{10}{9}y^4+x$}
For $y(0)=0$,
\begin{equation}
y_n'(0)\sim 1.1102(n+5.247)^{5/11}\quad{\rm and}\quad
y_n'(0)\sim -1.109(n-2.200)^{5/11}.\nonumber
\end{equation}
For $y'(0)=0$ and for negative $x$,
$$y_n(0)\sim1.80547(n-0.999)^{2/11}.$$
To leading order $\half\left[ y_n'(0)\right]^2\sim 0.6163\,n^{10/11}$ agrees
with $-\frac{2}{9}y_n^5(0)\sim 0.6155\,n^{10/11}$.
Again, with $y'(0)=0$ but for positive $x$,
$$y_n(0)\sim-1.226(n+0.152)^{2/11}.$$

\subsubsection{Generalized Painlev\'e 6a (GP6a): $y''=\frac{14}{25}y^6+xy^4$}
For $y(0)=0$ and for $x<0$,
\begin{eqnarray}
y_n'(0) &\sim& 3.06787\left\{\begin{array}{ll}(n+0.21)^{7/9} & (n\,{\rm odd}),\\
(n-0.27)^{7/9} & (n\,{\rm even}),\end{array}\right.\nonumber\\
y_n'(0) &\sim& -3.06786\left\{\begin{array}{ll}(n+0.73)^{7/9}& (n\,{\rm odd}),\\
(n-1.22)^{7/9} & (n\,{\rm even}).\end{array}\right.\nonumber
\end{eqnarray}
For $y(0)=0$ and for $x>0$,
\begin{equation}
y_n'(0)\sim2.9010(n-0.24)^{7/9}\quad{\rm and}\quad y_n'(0)\sim-2.9010(n+
0.24)^{7/9}.
\nonumber
\end{equation}

\subsubsection{Generalized Painlev\'e 6b (GP6b): $y''=\frac{14}{25}y^6+xy^3$}
For $y(0)=0$,
\begin{eqnarray}
y_n'(0) &\sim& 1.7408\left\{\begin{array}{ll}(n+0.05)^{7/11} & (n\,{\rm odd}),\\
(n+0.65)^{7/11} & (n\,{\rm even}),\end{array}\right.\nonumber\\
y_n'(0) &\sim& -1.7408\left\{\begin{array}{ll}(n-0.35)^{7/11} &(n\,{\rm odd}),\\
(n-1.05)^{7/11}, & (n\,{\rm even}).\end{array}\right.\nonumber
\end{eqnarray}
For $y'(0)=0$,
\begin{eqnarray}
y_n(0)&\sim&-1.52224\left\{\begin{array}{ll}(n+0.152)^{2/11} & (n\,{\rm odd}),\\
(n+0.852)^{2/11} & (n\,{\rm even}).\end{array}\right. 
\nonumber
\end{eqnarray}
To leading order $\half\left[y_n'(0)\right]^2\sim1.51519\,n^{14/11}$ is close to
$-\frac{2}{25}y_n^7(0)\sim 1.51521\,n^{14/11}$.

\subsubsection{Generalized Painlev\'e 6c (GP6c): $y''(x)=\frac{14}{25}y^6(x)
+xy^2(x)$}
For $y(0)=0$,
\begin{eqnarray}
y_n'(0) &\sim& 2.5979\left\{\begin{array}{ll}(n-0.1811)^{7/13}&(n\,{\rm odd}),\\
(n-0.2639)^{7/13} & (n\,{\rm even}),\end{array}\right.\nonumber\\
y_n'(0) &\sim& -2.598\left\{\begin{array}{ll}(n-0.735)^{7/13} &(n\,{\rm odd}),\\
(n-0.818)^{7/13} & (n\,{\rm even}).\end{array}\right.\nonumber
\end{eqnarray}
For $y'(0)=0$,
\begin{eqnarray}
y_n(0) &\sim& 1.73085 (n+0.098)^{2/13},\nonumber\\
y_n(0) &\sim& -1.7065\left\{\begin{array}{ll}(n-0.456)^{2/13} &(n\,{\rm odd}),\\
(n-0.54)^{2/13}, & (n\,{\rm even}).\end{array}\right.\nonumber
\end{eqnarray}
To leading order $\half\left[y_n'(0)\right]^2\sim 3.3745\,n^{14/13}$ is close to
$-\frac{2}{25}y_n^7(0)\sim 3.37158\,n^{14/13}$.

\subsubsection{Generalized Painlev\'e 6d (GP6d): $y''(x)=\frac{14}{25}y^6(x)+
xy(x)$}
For $y(0)=0$,
\begin{eqnarray}
y_n'(0) &\sim& 2.3569\left\{\begin{array}{ll}(n+0.591)^{7/15} &(n\,{\rm odd}),\\
(n+0.0632)^{7/15} & (n\,{\rm even}),\end{array} \right.\nonumber\\
y_n'(0) &\sim& -2.357\left\{\begin{array}{ll}(n-0.589)^{7/15} &(n\,{\rm odd}),\\
(n-0.0611)^{7/15} & (n\,{\rm even}).\end{array}\right.
\nonumber
\end{eqnarray}

\subsubsection{Generalized Painlev\'e 6e (GP6e): $y''(x)=\frac{14}{25}y^6(x)+x$}
For $y(0)=0$,
\begin{equation}
y_n'(0)\sim2.3219(n+0.36)^{7/17}\quad{\rm and}\quad
y_n'(0)\sim-2.322(n-1.04)^{7/17}.\nonumber
\end{equation}
For $y'(0)=0$,
\begin{eqnarray}
y_n(0) &\sim& 1.500998(n-0.9996)^{2/17},\nonumber\\
y_n(0) &\sim& -1.652812(n-0.3366)^{2/17}.\nonumber
\end{eqnarray}
To leading order,
$\half\left[y_n'(0)\right]^2\sim2.6956\,n^{14/17}$ agrees with $-\frac{2}{25}
y_n^7(0)\sim2.695592\,n^{14/17}$.

\subsubsection{Generalized Painlev\'e 7a (GP7a): $y''(x)=\frac{4}{9}y^7(x)+x
y^5(x)$}
For $y(0)=0$ and $x<0$,
\begin{eqnarray}
y_n'(0) &\sim& -1.86695\left\{\begin{array}{ll}
(n-0.1849)^{4/5} & (n\,{\rm odd}),\nonumber\\
(n-0.8144)^{4/5} & (n\,{\rm even}).\nonumber
\end{array}\right.
\end{eqnarray}
For $y(0)=0$ and $x>0$,
$$y_n'(0)\sim-2.29535(n-0.056)^{4/5}.$$

\subsubsection{Generalized Painlev\'e 7b (GP7b): $y''=\frac{4}{9}y^7+xy^4$}
For $y(0)=0$ and $x<0$,
$$y_n'(0)\sim-1.38115(n-0.2318)^{2/3}.$$ 
For $y(0)=0$ and $x>0$,
$$y_n'(0)\sim-1.38114(n-0.231)^{2/3}.$$

\section{Hyperfine splitting of eigenvalues}\label{s6}
In this section we study a new kind of eigenvalue problem for nonlinear
differential equations whose structure is analogous to the hyperfine spectrum in
atomic physics. This new kind of solution initially follows the usual $n$th
separatrix solution $Y_n(x)$. However it suddenly deviates from $Y_n(x)$, and
then undergoes $m$ rapid oscillations after which it approaches a different
limiting curve. We find that for each value of $n$ there are an infinte number
of choices for $m$. These new kinds of separatrix solutions are observed in GP4a
and GP6c.

We illustrate the hyperfine separatrix solutions using the GP4a equation:
\begin{equation}
y''(x)=\frac{1}{a^2}y^4(x)+xy^2(x)\quad{\rm with}\quad a=3/\sqrt{10}.
\label{E45}
\end{equation}
To obtain the hyperfine solutions we let
\begin{equation}
y(x)=Y_n(x)+\phi(x),
\label{E46}
\end{equation}
where $Y_n(x)$ is the $n$th conventional separatrix solution whose initial
slope is 0 and whose asymptotic behavior is
\begin{equation}
Y_n(x)\sim a\sqrt{-x}\quad(x\to-\infty).
\label{E46a}
\end{equation}
Note that for GP4a the initial value is negative, $Y_n(0)<0$.

\begin{figure}[h!]
\begin{center}
\includegraphics[width=0.8\columnwidth]{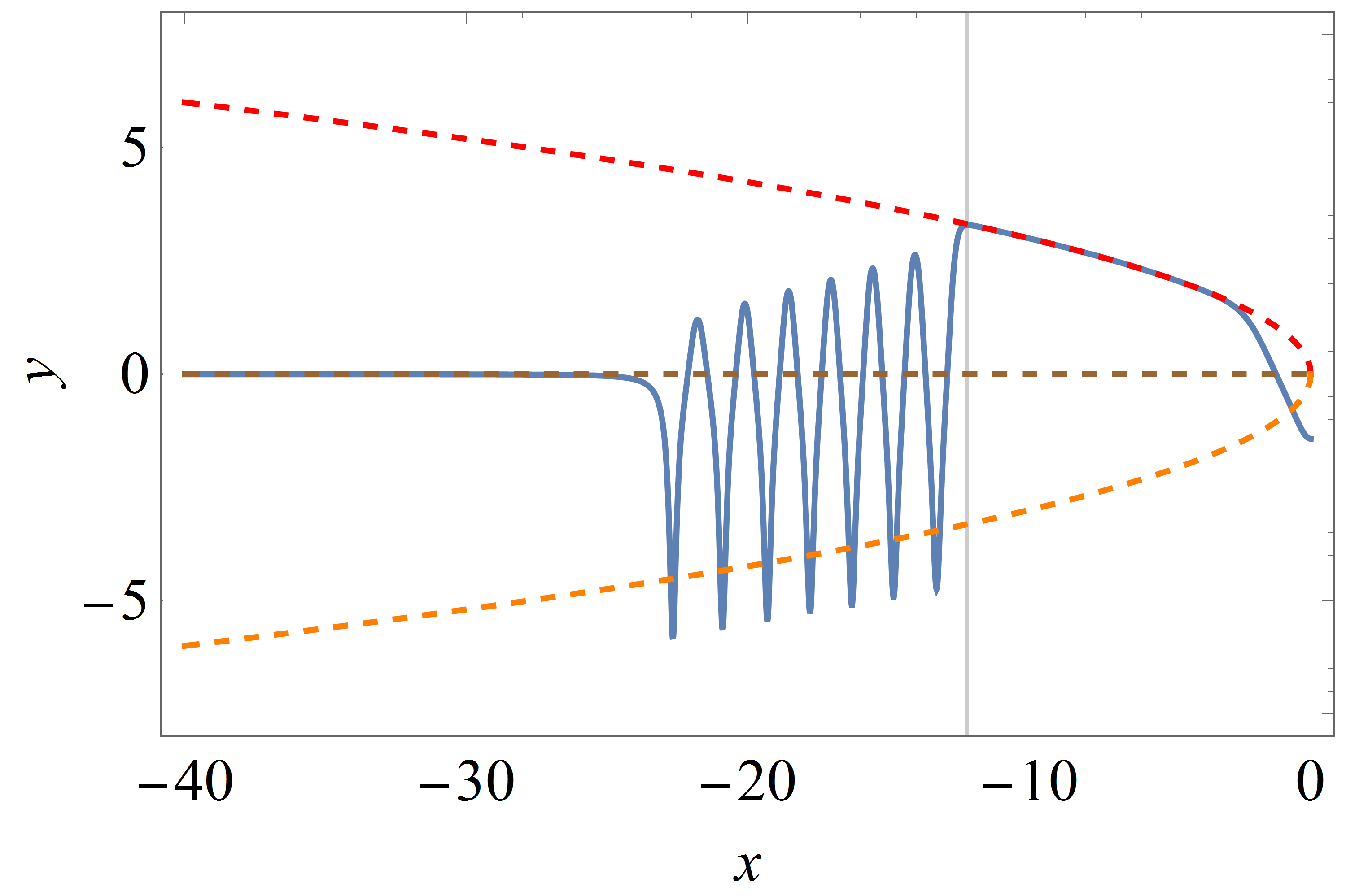}
\caption{\label{F4} Typical behavior of an eigenfunction of (\ref{E45}) that
exhibits hyperfine splitting. In this figure a conventional eigenfunction begins
at $x=0$ with vanishing slope and initial value $-1.427\,047\,040\,594\,516\,272
\,795\,825\,740$ (solid line to the right of the vertical line). As $x$ becomes
more negative, this eigenfunction follows the upper dashed curve $3\sqrt{-x/10
}$ all the way to $x=-\infty$. However, if we shift the initial value by the
extremely small amount $2.846\,993\,08\times 10^{-20}$, the hyperfine solution
departs from the conventional solution on the dashed curve near $x=-12.27$
(indicated by the vertical line) and undergoes seven rapid oscillations before
leveling off at the middle dashed curve on the negative-$x$ axis. There are an
infinite number of such hyperfine eigenfunctions with each successive
eigenfunction having one additional oscillation before leveling off at the
negative-$x$ axis. (See Table \ref{t1}.) The eigenfunction plotted here is the
seventh hyperfine eigenfunction associated with the $n=0$ conventional
separatrix solution.}
\end{center}
\end{figure}

We will see that the new hyperfine solution $y(x)$ initially follows the usual
separatrix solution $Y_n(x)$. However it deviates from $Y_n(x)$, and then
oscillates $m$ times about the negative-$x$ axis and the curve $-a\sqrt{-x}$.
Eventually, $y(x)$ approachies 0 like an inverse cubic:
$$y(x)\sim12x^{-3}\quad(x\to-\infty).$$
This behavior of $y(x)$ is shown in Fig.~\ref{F4} for the case $n=0$ and
$m=7$.

To derive the behavior shown in Fig.~\ref{F4}, we substitute $y(x)$ in
(\ref{E46}) into the generalized Painlev\'e equation (\ref{E45}). We see that
$\phi$ satisfies the nonlinear differential equation
\begin{equation}
\phi''=\left(\frac{4}{a^2}Y_n^3+2xY_n\right)\phi+\left(\frac{6}{a^2}
Y_n^2+x\right)\phi^2+\frac{4}{a^2}Y_n\phi^3+\frac{1}{a^2}\phi^4.
\label{E47}
\end{equation}
When does $y(x)$ stop following $Y_n(x)$? Before these curves separate,
$\phi(x)$ is small ($\phi\ll1$). Thus, we can approximate (\ref{E47}) by
the linear equation
\begin{equation}
\phi''\sim\left(\frac{4}{a^2}Y_n^3+2xY_n\right)\phi\sim2a(-x)^{3/2}\phi,
\label{E48}
\end{equation}
where we have neglected all higher powers of $\phi$ and we have substituted the
asymptotic behavior in (\ref{E46a}).

A straightforward WKB analysis of (\ref{E48}) shows that for large negative $x$
there are two possible asymptotic behaviors for $\phi$:
$$\phi(x)\sim\frac{C}{(-x)^{3/8}}\exp\left[\pm\frac{4}{7}\sqrt{2a}(-x)^{7/4}
\right].$$
For the GP4a eigenfunction problem for $Y_n(x)$ the initial values of $\phi$
are the hyperfine eigenvalues that we are seeking.

To illustrate the hyperfine behavior, we consider the lowest ($n=0$)
conventional separatrix eigenfunction $Y_0(x)$. An extremely precise numerical
study shows that for large $m$ the $m$th hyperfine eigenvalue is given
approximately by
$$\phi_m(0)\sim 4.1789\,e^{-9.262\,01 m}\quad(m\to\infty).$$
All $\phi_m(x)$ follow $Y_0(x)$ closely starting at $x=0$. Then depending on
each hyperfine eigensolution, $\phi_m(x)$ separates from $Y_0(x)$ near $x=-T_m<
0$. Since $\phi_m(x)$ and $\phi_{m+1}(x)$ nearly overlap when $T_m\leq x\leq0$,
we can assume that
\begin{equation}
\frac{\phi_{m+1}(0)}{\phi_m(0)}\sim\frac{\phi_{m+1}(-T_{m+1})}{\phi_m(-T_m)}.
\end{equation}
This leads to
$$e^{-9.26201}\sim\left(\frac{T_m}{T_{m+1}}\right)^{3/8}\exp\left[\pm\frac{4}{7}
\sqrt{2a}\left(T_{m+1}^{7/4}-T_m^{7/4}\right)\right].$$

The solution to this nonlinear recursion relation satisfies
$$9.262\,01m\sim\frac{4}{7}\sqrt{2a}\left(T_m^{7/4}-T_0^{7/4}\right)-\frac{3}{8}
\left(\log T_m-\log T_0\right)\quad(m\to\infty).$$
For large $m$, $T_m\gg T_0$ and $T_m^{7/4}\gg\log T_m$. Thus, we obtain the
leading asymptotic behavior of $T_m$:
\begin{equation}
T_m\sim\left(\frac{7}{4\sqrt{2a}}9.262\,01m\right)^{4/7}\quad(m\to\infty).
\label{E54}
\end{equation}
This formula fits the data strikingly well, as one can see in Table \ref{t1}.

\begin{center}
\begin{table}[htb]
\begin{center}
\begin{tabular}{clllllll}
\hline
\hline
$m$ & ~~~1  & ~~~2 & ~~~3 & ~~~4 & ~~~5 & ~~~6 & ~~~7 \\
\hline
$T_m$ &~$3.28$ & ~$5.66$ &~$7.35$ &~$8.78$ & $10.05$ & $11.20$& $12.27$ \\
\hline
Eq.~(\ref{E54})&  ~$4.09$ & ~$6.08$ & ~$7.66$ &~$9.03$ &$10.26$ &$11.39$ &  $12.44$ \\
\hline
\hline
$m$ & ~~~8 & ~~~9 & ~~10 & ~~11 & ~~12 & ~~13 &~~ 14\\
\hline
$T_m$      & $13.27$ &$14.22$ &$15.12$ &$15.99$ &$16.82$ & $17.61$ & $18.39$  \\
\hline\hline
Eq.~(\ref{E54})& $13.42$&$14.36$ &$15.25$ &$16.10$ &$16.92$ & $17.72$&
$18.48$ \\ 
\hline
\hline
\end{tabular}
\end{center}
\caption{\label{t1} Verification of the accuracy of (\ref{E54}). In this table
$T_m$ is the value of $-x$ at which the $m$th hyperfine eigenfunction splits
away from the zeroth conventional separatrix eigenfunction $Y_0(x)$ by $1\%$:
$\big[\phi(-T_m)-Y_0(-T_m)\big]/Y_0(-T_m)\approx1\%$. The theoretical
prediction is given by the right side of (\ref{E54}). Note that the accuracy
improves rapidly with increasing $m$.}
\end{table}
\end{center}

For GP4a there is only one class of hyperfine solutions. These solutions are
associated with the conventional initial-function separatrices. However, for
GP6c there are two classes of hyperfine eigenvalue solutions, one associated
with the initial-slope problem with $Y_n'(0)>0$ and another associated with
the initial-function problem $Y_n(0)>0$.

\section{Summary and future directions for research}
\label{s7}
This paper presents a broad study of eigenvalue phenomena associated with
nonlinear differential equations. The eigenfunctions are (unstable) separatrices
and the eigenvalues are the initial conditions that give rise to these
separatrix solutions. The large classes of nonlinear equations considered in
this paper have easily identifiable asymptotic behaviors for large $|x|$. In
some cases the large-eigenvalue behavior can be determined analytically by
reducing the nonlinear problem to the linear problem of finding the high-energy
eigenvalues of a quantum-mechanical Hamiltonian. However, such a linearization
procedure is only rarely possible to perform. We have presented a large array
of numerical studies, and we have also discovered the unexpected existence of
hyperfine eigenvalue structure.

\begin{figure}[h!]
\begin{center}
\includegraphics[width=0.8\columnwidth]{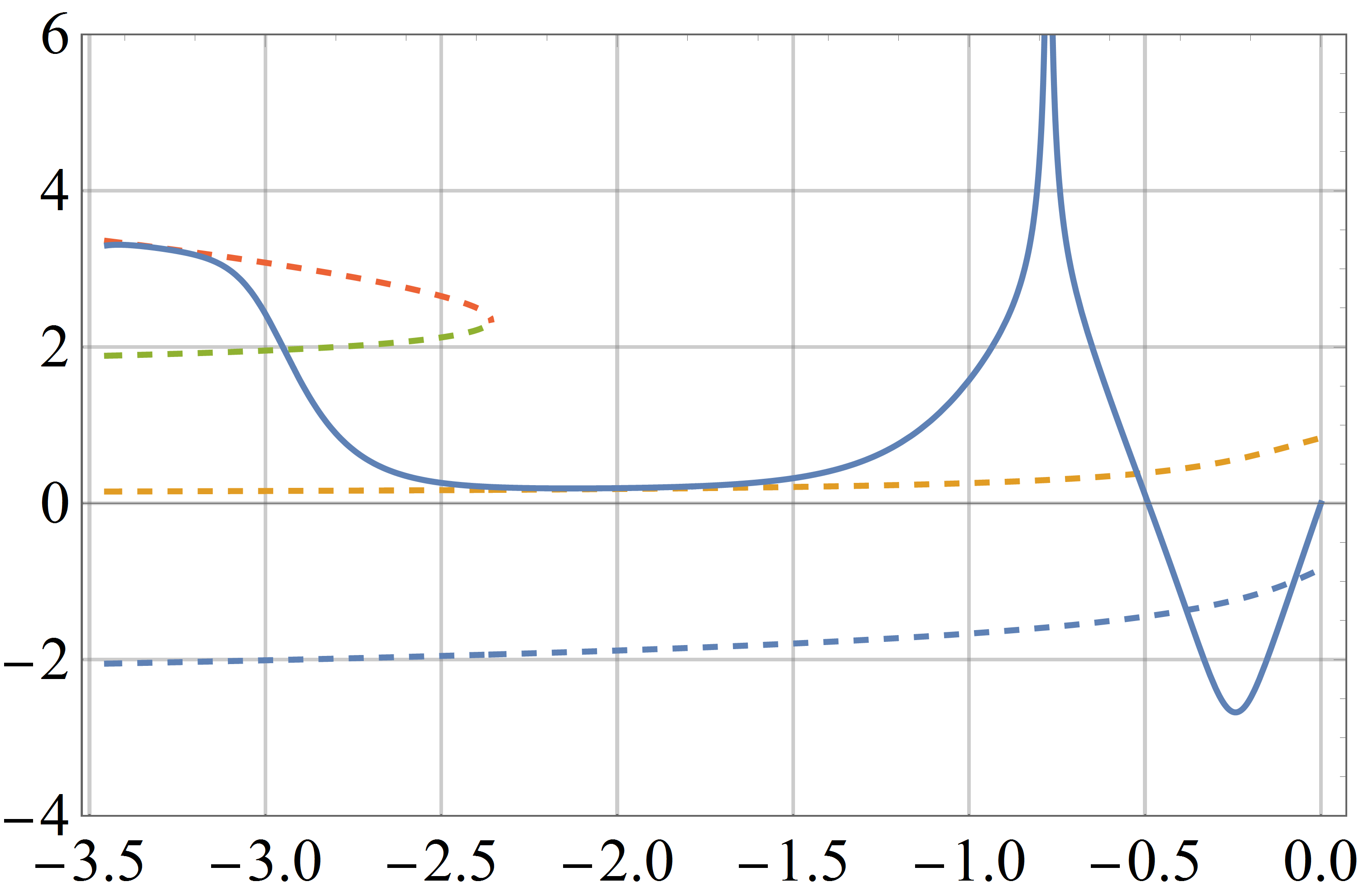}
\caption{\label{F5} Separatrix solution to the particular choice of generalized 
Painlev\'e equation in (\ref{E100}). In this figure, the dashed lines are the
fixed-point functions $f(x)$ that are obtained by setting the right side of
(\ref{E100}) to zero. The solid line represents a one-singularity
initial-slope separatrix eigensolution. This solution approaches the upper
unstable asymptotic behavior.}
\end{center}
\end{figure}

We conclude with a list of opportunities for future research on the general
topic of nonlinear eigenvalue problems. To begin, we emphasize that in this
paper we chose extremely simple forms for the polynomials $p(y)$ and $q(y)$ in
(\ref{E29}); specifically, we took $p(y)$ to be a {\it monomial} and we set
$q(y)=0$. Even with these elementary choices in (\ref{E35}) we have found a
rich set of eigenfunction behaviors, including hyperfine structure. The
possibilities for further study, both analytical and numerical, are immense. For
example, if we take $M=6$, take $p(y)$ to be a quartic polynomial, and take
$q(y)$ to be a cubic polynomial,
\begin{equation}
y''(x)=\textstyle{\frac{14}{25} y^6(x)+x \left[y^4(x)+5y^3(x)-\frac{2}{9}y^2
(x)-19y(x)+2\right]+\frac{5}{7}y^3(x)+4y^2(x)-\half y(x)-3},
\label{E100}
\end{equation}
we obtain eigenfunctions like that shown in Fig.~\ref{F5}. In this figure we see
{\it four} possible asymptotic behaviors and the separatrix eigenfunction that
is plotted approaches the uppermost (unstable) one of these behaviors.

A second avenue of research concerns the study of more general eigenvalue
problems. Until now, we have only studied initial-slope problems and
initial-function problems. It would be interesting to study mixed problems in
which both the initial function and the initial slope play the role of
eigenvalues, and it would be particularly interesting to study correlated
limits of initial-slope and initial-function problems.

Finally, we remark that we have only studied eigenvalue problems of the form
$y''(x)=F[x,y(x)]$ whose singularities are poles and algebraic branch points.
It would be of great interest to study the eigenfunctions assopciated with
nonlinear differential equations having logarithmic singularities, such as the
Thomas-Fermi equation. Furthermore, equations of the form
$y''(x)=F[x,y(x),y'(x)]$ should be studied, as well as higher-order-derivative
equations.

\acknowledgments
CMB thanks the Alexander von Humboldt Foundation for partial financial support.
QW thanks Sun Yat-sen University for its hospitality and Yu-Qiu Zhao for useful
discussions. QW is supported by the Singapore Ministry of Education Academic
Research Fund Tier I (WBS No.~R-144-000-352-112).


\begin{thebibliography}{99}

\bibitem{r1} C.~M.~Bender, A.~Fring, and J.~Komijani, J.~Phys.~A:
Math.~Theor.~{\bf 47}, 235204 (2014).

\bibitem{r2} C.~M.~Bender and J.~Komijani, J.~Phys.~A: Math.~Theor.~{\bf 48},
475202 (2015).

\bibitem{r3} C.~M.~Bender, J.~Komijani, and Q.-h.~Wang,
in {\it Resurgence, Physics and Numbers}, CRM (Centro di Ricerca Matematica)
Series, Ennio De Giorgi {\bf 20}, 67-90 (2017), ed. by F.~Fauvet, D.~Manchon,
S.~Marmi, and D.~Sauzin.

\bibitem{r1a} O.~S.~Kerr, J.~Phys.~A: Math.~Theor.~{\bf 47}, 368001 (2014).

\bibitem{R1} E.~L.~Ince, {\it Ordinary Differential Equations} (Dover, New
York, 1956).

\bibitem{R2} J.~W.~Miles, Proc.~Royal~Soc.~London A {\bf 361}, 277 (1978).
\bibitem{R3} P.~Holmes and D.~Spence, Quart. J. Mech.~Appl.~Math.~{\bf 37},
525 (1984).

\bibitem{R4} S.~P.~Hastings and J.~B.~McLeod, {\it Classical methods in
ordinary differential equations: With applications to boundary value problems},
Graduate Studies in Math. {\bf 129} (American Mathematical Society, 2011).

\bibitem{R5} Asymptotic behavior of the Painlev\'e transcendents is discussed
in M.~Jimbo and T.~Miwa, Physica D {\bf 2}, 407 (1981). 

\bibitem{R6} Separatrix behavior of the first Painlev\'e transcendent is
examined briefly in A.~A.~Kapaev, Differential Equations {\bf 24}, 1107 (1989);
see also A.~A.~Kapaev, CRM Proc.~Lect.~Notes {\bf 32}, 157 (2002).

\bibitem{R7} P. A. Clarkson, J.~Comp.~Appl.~Math.~{\bf 153}, 127 (2003).

\bibitem{R8} D.~Maseoro, {\it Essays on the Painlev\'e First Equation and the
Cubic Oscillator}, PhD Thesis, SISSA (2010).

\bibitem{R9} T.~Kawai and Y.~Takei, {\it Algebraic Analysis of Singular
Perturbation Theory}, (American Mathematical Society, New York, 2005).

\bibitem{R10} O.~Costin, R. D. Costin, and M. Huang, {\it Tronqu\'ee solutions
of the Painlev\'e equation $P_1$} (2013, unpublished).

\bibitem{R11} A.~S.~Fokas, A.~R.~Its, A.~A.~Kapaev, and V.~Y.~Novokshonov,
{\it Painlev\'e Transcendents: The Riemann-Hilbert Approach},
(American Mathematical Society, New York, 2006).

\bibitem{r4} C.~M.~Bender and S.~Boettcher, Phys.~Rev.~Lett.~{\bf 80}, 5243
5246 (1998).

\bibitem{r5} W.-G.~Long, Y.-T.~Li, S.-Y.~Liu, Y.-Q.~Zhao,
Stud.~App.~Math.~{\bf 139}, 505-532 (2017).

\bibitem{r6} C.~M.~Bender and S.~A.~Orszag, {\it Advanced Mathematical Methods
for Scientists and Engineers} (McGraw Hill, New York, 1978).

\end{thebibliography}
\end{document}